\documentstyle[aps,pra,epsf,floats,floats]{revtex}

\newcommand{\be}{\begin{equation}}
\newcommand{\bfig}{\begin{figure}}
\newcommand{\ee}{\end{equation}}
\newcommand{\efig}{\end{figure}}
\newcommand{\bi}{\begin{itemize}}
\newcommand{\ei}{\end{itemize}}
\newcommand{\bear}{\begin{eqnarray}}
\newcommand{\eear}{\end{eqnarray}}
\newcommand{\ba}{\begin{array}}

\newcommand{\ea}{\end{array}}

\newcommand{\bs}{\begin{slide}}
\newcommand{\es}{\end{slide}}
\newcommand{\bc}{\begin{center}}
\newcommand{\ec}{\end{center}}

\begin{document}
\ifpreprintsty \else
\twocolumn[\hsize\textwidth\columnwidth\hsize\csname@twocolumnfalse%
\endcsname \fi
\draft

\title {Criticality in a  model of banking crises.}
\author{Giulia  Iori }
\address{ Department of Mathematics, Kings
College \\ Strand, London WC2R 2LS, U.K. \\ E-mail:
giulia.iori@kcl.ac.uk}

\author{Saqib Jafarey}
\address{Department of Economics,
University of Wales Swansea,\\ Singleton Park, Swansea  SA2 8PP,
U.K.\\ E-mail: s.s.jafarey@swansea.ac.uk}

\maketitle

\vskip .5cm \centerline{(\today)}

\begin{abstract}

An interbank market lets
participants pool the risk arising from the combination of 
illiquid investments and random withdrawals by depositors. 
But it also creates the potential for
one bank's failure to trigger off  avalanches of further failures.
We simulate a model of interbank lending to study the interplay of 
these two effects.
We show that when banks are similar in size and exposure to risk, 
avalanche effects are small 
so that widening the interbank market leads to more stability.  
But as heterogeneity increases, avalanche effects become more 
important. By varying the heterogeneity and connectivity across banks, 
the system enters a critical regime with a power law 
distribution of avalanche sizes. 

\end{abstract}

\ifpreprintsty \else
] \fi              

\section{Introduction}

Systemic failure in banking arises when one bank's  failure 
triggers off further failures. The history of 
modern banking is full of examples of systemic failure 
at both moderate and large scales, with the 1997 East Asian 
crisis being 
the most recent example of large scale bank failure. 

Several channels through  
which systemic failure may arise have been explored 
\cite{AMR,CK,DD,ELM,F,JB,RT}. In this paper, we 
focus on one
channel, namely interbank lending. Interbank lending 
allows banks which face temporary shortfalls in funds to borrow 
from banks which have surpluses. Such 
 lending creates a network of credit and debt relationships within
the banking system. If one debtor bank fails, it 
adversely affects the balance sheet of its creditors 
 and can trigger off their subsequent failures. 

At the same time, interbank lending allows for the pooling of risk 
arising from the stochastic pattern of deposits and withdrawals by   
each bank's customers. Without interbank lending,
banks might become even more vulnerable to 
failure, although such failure would be idiosyncratic and 
free of the symptoms of systemic collapse. Judgment on the effects of
interbank lending for a banking system's stability should therefore take
into account both the {\em ex ante} risk-sharing effect and the {\em ex
post} systemic failure effect. In this paper, we study 
these effects by developing and simulating a dynamic
model of a  banking system, linked together 
through an interbank credit market.      

Our approach
abstracts from the
rationality of individual behavior and resembles the
approach taken in the statistical mechanics of disordered systems. 
Episodes of systemic failure in banking appear analogous to 
the periodic collapses or `avalanches' which arise in 
natural systems in the form of
earthquakes, microfracturing, epidemics, magnetization of ferro-magnetic
systems, etc.  The capacity of a
natural 
system to
generate avalanches of various sizes seems to depend on certain general
features, 
such as the heterogeneity of the elementary constituents of a system and
the nature of interaction between them. By analogy in a banking system, 
we focus on how the individual characteristics of banks  
 and their mutual interaction on an interbank
market can affect the stability of the system.

\section{ The Model.}

Time is discrete. At the initial time, $t=0$, the system starts with 
$N_0$ banks.  At any subsequent time,  there are $N_t$ banks operating
in the system. The number of banks may go down as a result of failures,
but failing banks are not replaced by new entrants. 

 Credit linkages between banks are 
 defined by a connectivity 
matrix, $J_{ij}$.  $J_{ij}$ is either one or zero; a value 
of one indicates that a credit linkage exists
 between banks $i$ and $j$ and zero indicates no relationship. $J_{ij}$
 are randomly chosen at the beginning of the simulation. $c$ denotes the 
 probability that $J_{ij}$ is one for any two banks. 
At one extreme, $c=0$
 represents the case of no interbank lending, while $c=1$ represents a
 situation in which all banks can potentially borrow and lend from each 
 other.

The primary purpose 
of a bank is to channel funds received from depositors towards productive
investment. 
Deposits,  $A^{k}_t$, and  investment opportunities,
$\omega^{k}_t$,
 are randomly and independently drawn for each bank $k$ at each time $t$
from  Gaussian distributions with means $\bar A^k$ and $\bar \omega^k$
respectively,   and variances  $\bar A^k \sigma_A$ and $\bar \omega^k \sigma_\omega$ respectively: 
$$
A^k_t = |\bar A^k  (1 + \sigma_A  \epsilon^1_t)| 
$$
$$
\omega^k_t =  |\bar \omega^k   (1 + \sigma_{\omega}  \epsilon^2_t) |
$$
$$
\epsilon^{1,2} \sim N(0,1)
$$

 Once tied up in investment,
resources become relatively illiquid; any investment made
at time $t$ fully matures only at some time $t+\tau$. 
A return, $\rho$, is realized at each point, $t+1$, $t+2$,
$\cdots$ $t + \tau$. $\rho$ is exogenous and risk-free. However, since 
withdrawals are unpredictable, a
bank may find it itself unable to repay its depositors due to the
illiquidity of its investment. 

Each bank $k$ starts a period, $t$, with cash holdings inherited from 
the past, $M^k_{t-1}$. These are constituted by    
$$
 M^k_{t-1} =  A^{k}_{t-1} + B^{k}_{t-1} +
 V^{k}_{t-1} - \sum_{s=1}^{\tau} I^{k}_{t-s},
$$
where $A^k$ are its deposits, $V^{k}$ represents its  equity, $I^{k}$ represents
past investments  and $B^{k}$ represents borrowings. 
$B^k$ can be negative or positive and satisfies: $\sum_{k=1}^{N} B^k
= 0$. Borrowing consists of one-period loans,
 which require repayment in full in the period after which they are
 undertaken.

Each bank receives income, $\rho \sum_{s=1}^{\tau-1}
I^k_{t-s} $ and $(1 +\rho )I^k_{t-\tau}$ from investments made
over the last $\tau$ periods.  It then receives
a new level of $A^k_t$, as defined above. Its cash holdings then adjust
to reflect $(A^{k}_t - A^k_{t-1})$, the net change in its deposits 
and $\rho\sum_{s=1}^{\tau}I^k_{t -s} + I^k_{t-\tau}$, the receipt of
incomes from past investment. If the adjusted cash holdings become
negative,  the bank can issue negotiable
debt certificates to cover any excess of payments over its cash
reserves. However, the certificates have to be redeemed at the
end of the period through borrowing from other banks. If this is
not done, the bank fails and its debt certificates become
worthless.

If a bank 
has borrowed in the past period, its priority is to repay
its creditors. Such payments are 
 made in cash. Hence, 
 banks
 with cash holdings in excess of their debt obligations,
which equal $(1+r_b)B^{j}_{t-1}$ for bank $j$,  make payments 
to their respective creditors.
 Here, $r_b$ is the  exogenous 
 interest rate on interbank borrowing. 

 At this point, two types of banks can be
distinguished, those with positive cash and those with negative
cash.  Accordingly,
 they get classified as  potential lenders and potential
 borrowers respectively. Borrowing banks issue demands for loans equaling their debt
 obligations (interest plus principal) minus their current cash
 (which if negative adds to their demand).

Lending banks give priority to dividend payments (to shareholders) 
 and investment. Dividend payments are made for  
 maintaining a target equity:deposit ratio, $\chi$, 
 and for preventing excessive 
levels of capitalization. Only  banks
 whose adjusted cash holdings plus illiquid assets exceed a certain
 fraction of their total deposits make dividend payments.   
 After dividends have been paid, the bank is
assumed to undertake  investment on the basis of its available
liquid resources on the one hand, and its stochastic investment
opportunity on the other. Available liquid resources comprise the
 bank's current cash minus any statutory reserve requirement imposed on
 it by a regulator. If the available investment opportunity exceeds this 
limit, the bank cannot exploit the full opportunity. It then invests
 upto the extent of its available cash resources.

 After investment, any excess left over is made available
 to those borrowing banks with whom a credit linkage 
exists.    Each borrowing bank contacts each lending bank with which it
 is linked in a random order.  The two banks exchange an amount of credit
equal to the 
minimum of the two banks' respective demand and supply. If the borrowing bank
 is left with an unfulfilled trade it contacts
another lender with which it is linked.

 A borrowing bank does not receive actual funds
until it has lined up enough credit to ensure that it will not
fail during the current period. Once a bank has obtained
sufficient credit, funds are transferred and the cash positions
of all banks involved is updated. This continues until either all
loanable funds are exhausted or all demands for credit are
satisfied.

 At this point, the process reiterates itself
through the following steps: banks which had not repaid creditors
in the first round but now have borrowed enough cash to pay off
past debt entirely, do so; these payments go to their creditors
and money holdings get updated accordingly; potential borrowers
and lenders are determined for the next round; if lenders had 
underutilized investment opportunities from the previous round, they 
make further investment. Finally, a fresh round of
borrowing and lending takes place. The process
repeats itself until reiteration produces no further exchange of
credit.

  All banks which are left with negative cash
holdings or cash holdings which fall short of their remaining
debt obligations are deemed to be in default. These banks are
removed from the system.
If, at the time of closure,  a failing bank  has illiquid assets
in the form of investments made previously, these are 
liquidated at a fraction, $\gamma$ (which is exogenous), of their true
value. The proceeds are  distributed,  first to the depositors,
then, if there
is still some value
 left, this goes  to creditors from the previous period and  finally to
the shareholders.

\section{Simulations and Results.}

 In all the simulations to be  presented, the return
on investment, $\rho$, was risk-free and equal to $1.0 \%$ ,
the interest rate on interbank borrowing, $r_b$, was $0.5$, 
$\chi$, the equity:deposit ratio was $30 \%$, 
the period of maturity
$\tau$ was set at 3 and the recovery rate $\gamma$ was set at 0.
Parameters which
 vary across the reported simulations are identified
separately.

At $t=0$ each bank holds deposits,
$A_{0}$, worth 1000 units and initial equity, $V_{0}$, equal to
$0.3$ times the initial deposit.

Initially, $\bar A_k$ and $\bar \omega_k$ were chosen as identical 
for each bank. This led to homogeneity in average size of deposits 
 and investment
opportunities across banks. 

Figure 1 displays results on the main question
of the paper. It compares bank failures with different
degrees of linkage, $c$,  in the interbank market.
 Increasing linkage adds
stability, in the sense that in any period, there are more
surviving banks the greater the degree of linkage. This pattern
was very robust to changing the parameters of the simulation.

A policy issue raised by  the results of Figure 1  
concerns the role 
of statutory reserve requirements. By 
preventing  banks from investing  more than a certain fraction
of the deposits placed by customers in the form of illiquid and/or risky
assets, reserve requirements reduce the exposure of individual 
banks to the risk of failure. But they also  inhibit interbank lending
activity. This can reduce the risk sharing provided by interbank credit 
and destabilise the system. We
conducted experiments along the lines of those reported in Figure 1 and
found that, indeed, while without any interbank credit higher reserve requirements
always led to fewer bank failures,  with interbank credit linkages, 
 similar increases in reserve requirements could increase the incidence
of bank failures (see \cite{IJ}).

In the homogeneous case depicted in Figure 1, episodes of bank failure 
affect only a 
small number of banks at a time and there appears to be no correlation 
across the episodes. This is because there is relatively little borrowing 
and lending taking place at any time.  Even this small amount appears to
save banks which face short run liquidity problems, but is not enough to
generate contagion effects.

\begin{figure}[h]
\centerline{\epsfxsize 7cm  \epsffile{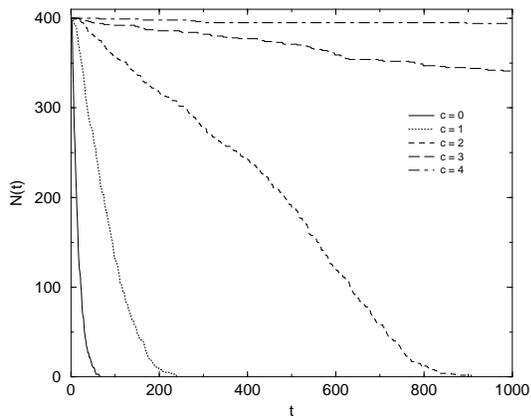}}
\caption{Surviving banks with different interbank linkages.}
\end{figure}

To allow for systematically larger volumes of interbank activity, we 
try to differentiate banks in a way that some banks tend to 
be lenders and others to be borrowers. We therefore 
 make  banks differ according to investment opportunities, 
by choosing:
$ \bar \omega^k = \bar \omega |z^k|$ 
 with  $z^k \sim N(0, \sigma_{\bar \omega})$. 
Given that banks remained identical in terms 
of their  deposit fluctuations, a large $\bar \omega_k$ meant that a bank
would invest large amounts and consequently face a greater exposure to
liquidity risk. Such a bank would be more likely to act as a
large borrower on the credit market, while for analogous reasons, a bank
with small $\bar \omega_k$ would be more likely to act as a large lender.

\begin{figure}[h]
 \centerline{  \epsfxsize 7cm \epsffile{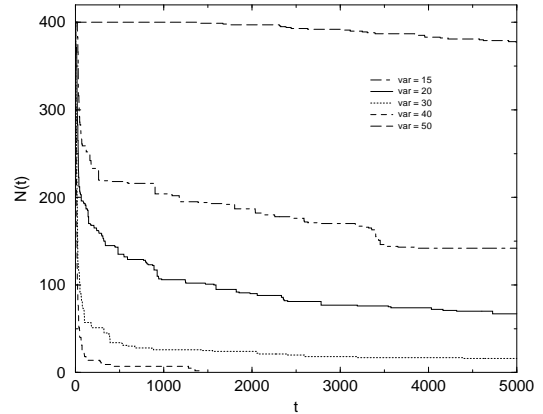}}
\caption{Effect of variance in opportunities on failure: 100
percent linkage.}

\end{figure}

Figures 2 and 3 show some of the results from
this experiment. 
 Figure 2 shows the effect of greater
heterogeneity on the incidence of bank failures when banks are
100 percent connected to each other. As the variance of types
increases from 5 to 40, the incidence of failures goes up. Note
that now the time path shows `avalanches', {\em i.e.} many banks
collapse together over brief periods of time. Since the aggregate
environment is on average constant, avalanches suggest that
knock-on effects are taking place.

 Figure 3  shows that, fixing the variance of types  at 10,
increasing connectivity from 5 to 50 percent leads to less
failures during the interval studied, but increasing connectivity
to 80 percent leads to more failures than at 50 percent. Hence,
in contrast to the case of homogeneous banks, increasing the
extent of the interbank market need not increase overall
stability.

\begin{figure}[h]
   \centerline{ \epsfxsize 7cm \epsffile{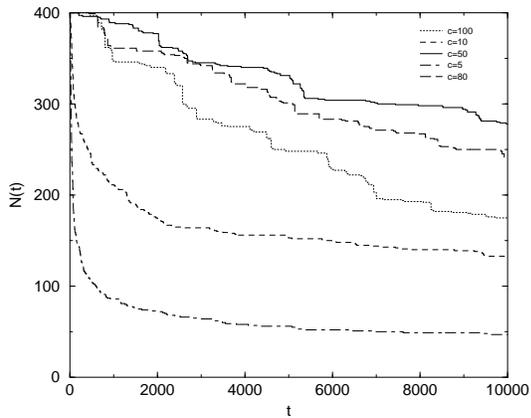}}
\caption{Effect of increasing linkages on failure; heterogeneous
investment opportunities. }

\end{figure}

We repeated the experiments discussed in Figures 2 and 3 by
differentiating banks both in terms of size and in terms of investment 
opportunity, as follows:
$\bar A^k   =  \bar A |z^k|$, and 
$\bar \omega^k  =  \bar \omega |z^k|$,
with $z^k \sim N(0, \sigma_{\bar \omega})$. With this formulation, all
banks face the same liquidity risk, but heterogeneity exists 
in the volume of borrowing and lending by different banks. A big
borrowing bank might need to contact many small lending banks and a big
lending bank might be a source of liquidity to many small banks. In 
both cases, the collapse of a big bank can create ripples 
through the system. The simulation results verified that this 
indeed does happen \cite{IJ}. These results are also consistent with empirical 
observations made in the literature (see, {\em e.g.}, \cite{AMR,F}). 

The results suggest that when banks are heterogeneous in the
volume of their activity on the interbank market, 
contagion effects can arise which can undermine the
overall risk sharing influence of interbank credit.   It is 
interesting to observe that heterogeneity 
plays a similar role in 
a well known physical model, the  Random Field Ising Model
(RFIM). In particular, the 
RFIM 
shows a 'disorder' induced phase transition between a regime
displaying only finite size avalanches and a regime corresponding
to avalanches which cover the whole system. At a critical level
of disorder, avalanches of all sizes, up to and including the
system as a whole, may occur. The statistical distribution of
avalanches takes the form of a power-law at the critical value \cite{SDK}.

\begin{figure}[h]
\centerline{  \epsfxsize 7cm  \epsffile{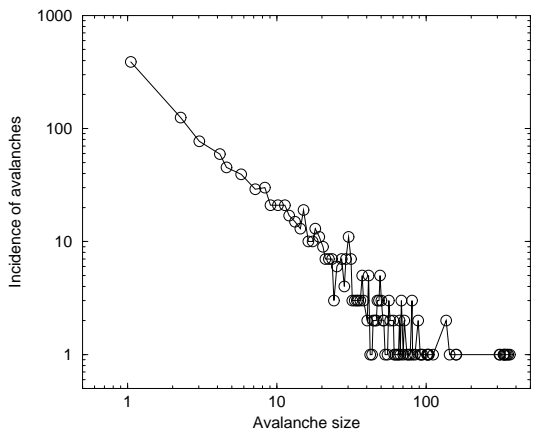}}
\caption{Log-log plot of the statistical distribution of
avalanche size. }

\end{figure}

A  similar phase transition is found in our model. 
For example, Figure 4 shows that when  linkages are fixed at 100 
percent and setting $\sigma_{\bar \omega} = 40$,  the system
reaches a critical state where the size of avalanches becomes
distributed according to a power-law. In order to
better identify this power-law behaviour, we increased the number
of banks to 1600 and repeated the simulations several times,
resampling  the investment opportunities each time.
A preliminary investigation indicates that for each value of
linkage across banks, it is possible to find a critical value of
heterogeneity in the system which leads to power-law
distributions of  avalanches sizes. The variance of sizes needed
to attain the power-law distribution  increases as the connectivity
decreases.

\section{Conclusions:}

Our simulations have identified certain characteristics which 
can lead interbank lending  
to be associated with complex forms of instability in 
a banking system. Whether
or not the crises which periodically grip actual banking systems 
represent such complex behaviour is difficult 
to establish empirically. The data needs for such measurement 
outstrip the recorded evidence on the subject.  It is nonetheless 
useful for policy makers to be able to distinguish circumstances 
under which interbank credit can destabilise a system 
from those under which it stabilises. In this respect, our simulations 
suggest that when bank lending is extended across banks which are 
heterogeneous in size and risk exposure, it can exarcebate the 
potential for avalanche effects and destabilise the system.


\begin{thebibliography}{99}

\bibitem{AMR}
 Angelini, P., G.~Maresca and D.~Russo (1996),
Systemic risk in the netting system, {\em Journal of Banking and
Finance}, 20, 853.

\bibitem{CK}
Calomiris, C.W. and C.M.~Kahn (1996), The
efficiency of self-regulated payment systems: Learning from the
Suffolk system, {\em Journal of Money, Credit and Banking}, 28,
766.


\bibitem{DD} Diamond, D.W.  and P.H.~Dybvig T. (1983), Bank
runs, deposit insurance and liquidity, {\em Journal of Political
Economy}, 91, 401.


\bibitem{ELM} Edison, H.J.,  P. Luangaram and M. Miller (2000),
Asset bubbles, domino effects and "lifeboats": Elements of the
East Asian crisis',  {\em Economic Journal}, 110.


\bibitem{F}  Furfine, C. (1999), Interbank exposures:
Quantifying the risk of contagion, {\em mimeo}, Bank for
International Settlements.

\bibitem{JB} Jacklin, C.J. and S.~Bhattacharya (1988),
Distinguishing panics and information-based bank runs: Welfare
and policy implications, {\em Journal of Political Economy}, 96,
568.

\bibitem{RT} Rochet, J-C and J.~Tirole (1996), Interbank
lending and systemic risk, {\em Journal of Money, Credit and
Banking}, 28, 733.

\bibitem{IJ} Iori, G. and S. Jafarey (2000), Interbank lending,
 reserve requirements and systemic risk, King's College preprint KCL-MTH-0060.

\bibitem{SDK}
Sethna, J.P., K.A. Dahmen, S. Kartha, J.A. Krumhansl,
 B. W. Roberts, and J.D. Shore  (1993), {\em Physical Review Letters},
 70, 3347.

\end{thebibliography}
\end{document}